\documentclass[conference]{IEEEtran}
\IEEEoverridecommandlockouts
\usepackage{cite}
\usepackage{amsmath,amssymb,amsfonts}
\usepackage{algorithm, algpseudocode}
\usepackage{graphicx}
\usepackage{textcomp}
\usepackage{balance}
\usepackage{xcolor}
\usepackage{graphicx}
\usepackage{placeins}
\usepackage{multirow}
\usepackage{subcaption}
\def\BibTeX{{\rm B\kern-.05em{\sc i\kern-.025em b}\kern-.08em
    T\kern-.1667em\lower.7ex\hbox{E}\kern-.125emX}}
\begin{document}


\title{Automated and Systematic Digital Twins Testing for Industrial Processes
}

\author{\IEEEauthorblockN{Yunpeng Ma}
\IEEEauthorblockA{\textit{Dep. Mathematics \& Computer Science } \\
\textit{Karlstad University}\\
Karlstad, Sweden \\
yunpeng.ma@kau.se}
\and
\IEEEauthorblockN{Khalil Younis}
\IEEEauthorblockA{\textit{Dep. Mathematics \& Computer Science} \\
\textit{Karlstad University}\\
Karlstad, Sweden \\
khalilyounis20@gmail.com}
\and
\IEEEauthorblockN{Bestoun S. Ahmed}
\IEEEauthorblockA{\textit{Dep. Mathematics \& Computer Science} \\
\textit{Karlstad University}\\
Karlstad, Sweden \\
bestoun@kau.se}
\and
\IEEEauthorblockN{Andreas Kassler}
\IEEEauthorblockA{\textit{Dep. Mathematics \& Computer Science} \\
\textit{Karlstad University}\\
Karlstad, Sweden \\
andreas.kassler@kau.se}
\and
\IEEEauthorblockN{Pavel Krakhmalev}
\IEEEauthorblockA{\textit{Dep. Materials Science} \\
\textit{Karlstad University}\\
Karlstad, Sweden \\
pavel.krakhmalev@kau.se}
\and
\IEEEauthorblockN{Andreas Thore}
\IEEEauthorblockA{\textit{Smart Industrial Automation} \\
\textit{RISE Research Institutes of Sweden}\\
Västerås, Sweden \\
andreas.thore@ri.se}

\and
\IEEEauthorblockN{Hans Lindbäck}
\IEEEauthorblockA{\centerline{\textit{Bharat Forge Kilsta AB}} \\
Karlskoga, Sweden\\
Hans.Lindback@bfkilsta.com}
}

\maketitle

\begin{abstract}
Digital twins (DT) of industrial processes have become increasingly important. They aim to digitally represent the physical world to help evaluate, optimize, and predict physical processes and behaviors. Therefore, DT is a vital tool to improve production automation through digitalization and becomes more sophisticated due to rapidly evolving simulation and modeling capabilities, integration of IoT sensors with DT, and high-capacity cloud/edge computing infrastructure. However, the fidelity and reliability of DT software are essential to represent the physical world. This paper shows an automated and systematic test architecture for DT that correlates DT states with real-time sensor data from a production line in the forging industry. Our evaluation shows that the architecture can significantly accelerate the automatic DT testing process and improve its reliability. A systematic online DT testing method can significantly detect the performance shift and continuously improve the DT's fidelity. The snapshot creation methodology and testing agent architecture can be an inspiration and can be generally applicable to other industrial processes that use DT to generalize their automated testing.
\end{abstract}

\begin{IEEEkeywords}
Digital twin, software testing, reinforcement learning, industry 4.0, machine learning
\end{IEEEkeywords}


\section{Introduction}
Over the past two decades, many companies have faced production problems that have led to low efficiency, poor product quality, and high costs. In particular, traditional and heavy industries, including steel and metal processing, also face material waste challenges that have a significant environmental impact and reduce profits. An essential reason for product defects and high cost is the low degree of automation and digitalization of the existing equipment used on the shop floor \cite{Cagle2020Digitalization}. For example, a forging factory typically relies on old heavy machinery, such as large induction ovens that are used to heat steel bars. The heating process of a forging line is essential for the final product quality; however, it is still manually controlled or by using well-known recipes (predefined parameter settings from experience) created by experts and experienced operators. The standard recipe may fail when unseen events occur. Here, an inaccurate manual temperature adjustment to heat the steel bars may not meet the required production requirement. Therefore, poor process control can lead to material quality degradation or even waste because the production temperature is outside the specification and the quality of the product cannot be automatically detected during production. Many companies are investing more resources to improve digitalization and automation in process control to overcome this problem.  With the improvement of digitalization, more key enabling technologies, such as artificial intelligence (AI) and industrial Internet of Things (I-IoT), can be integrated into the company's equipment to achieve AI-assisted automation\cite{ZDRAVKOVIC21AI-DHS}. As one of the emerging technologies, digital twins (DT) are becoming increasingly important in improving digitalization in the process industry due to rapidly evolving simulation and modeling capabilities \cite{Matteo2022DT}, also using the vast amount of compute processing from edge/cloud infrastructures.

A DT is a high-fidelity virtual model aimed at emulating physical processes and behaviors with the ability to evaluate, optimize, and predict\cite{Graessler2017DT}. It also provides a real-time link between the physical and digital worlds, allowing companies to have a complete digital footprint of their products and production processes throughout the production cycle \cite{Khan2018DT}. In turn, companies can realize significant value in improved operations, reduced defects, increased efficiency, and enabling predictive maintenance \cite{Panagou2022PM}. With real-time data acquisition, a DT can help operators understand the production process and make preventive decisions when an anomaly occurs \cite{LIU2021DT}. However, to make the correct decisions, DTs need to accurately simulate the physical world, which is difficult due to the complexity involved. Consequently, DTs representing the production process must be systematically tested to prove their reliability.

Testing DT as a software product is important because it identifies DT imperfections and helps increase its precision, ensuring high performance and reliable operation. With the acquisition of a large amount of production data, providing a continuous automated testing approach is crucial to ensure the overall quality of DT software. This testing process becomes more complicated when it should be done in real-time by combining the DT with the streaming production data. This is challenging because DT-based data processing is complicated due to the complexity of the physical process and the lack of consistency in the data stream. Most of the literature focuses on the use of DT to test different applications, where the DT itself is still manually tuned and tweaked with offline experimental data. Less evidence can be found for DT validation with online data streams. However, the performance of the well-tuned DT with offline data may change when the production environment changes. Hence, continuously monitoring the DT deviation from the real production line and continually building an ultra-fidelity virtual replica by entirely using physical live data streams is essential.

To address the mentioned challenges, in this paper, we provide a systematic and automated real-time method to continuously test the quality of the DT software with live production data in the forging factory (Bharat Forge Kilsta AB, Karlskoga, Sweden). This is a significant contribution as it allows the DT to be tested and validated using live production data rather than relying on simulated or historical data. The method is essential in industrial process automation, ensuring that the DT is accurate and reliable when used to optimize and control the process. The paper also presents a snapshot creation methodology that allows the DT to be tested in a systematic and automated way. The snapshot method allows the DT to be continuously tested and monitored, ensuring that it is always up-to-date and accurate. To discover faulty data, a snapshot creation methodology repetitively creates two snapshots, feeds the first snapshot to the DT, and compares the output of the DT with the second snapshot. The method identifies the deficiencies between the two snapshots and localizes the faults in the DT data. The paper also presents an architectural implementation of the testing method within the DT when a machine learning (ML) solution is used for the power control optimization algorithm.

To this end, the contributions of this paper are as follows:  we 1) propose a DT test architecture for real-time data parsing, processing, visualization, and analysis, 2) propose a novel method to test DT with real production data in real-time, and 3) provide a stable and scalable approach to processing the production data. To address these contributions, this paper is organized as follows. In Section \ref{sec:RelatedWork}, we provide the necessary background and related work. Section \ref{sec:DTtest} presents our case study in which the heating line of a forging factory, the functionality of the DT, and the DT-assisted DRL structure will be explained in detail. Section \ref{sec: DT_Architecture} illustrates the DT testing architecture and the snapshot creation methodology. Section \ref{sec:results} contains the experimental evaluation, including the testing setup, the experimental results, and the discussion. Section \ref{ThreatsToValidity} explains the threats to validity, generalizability, and the applicability of the proposed method. Finally, Section \ref{sec:conclusion} concludes the paper.


\section{Background and Related Work}\label{sec:RelatedWork}

In recent years, relevant academics have investigated and evaluated the concept, reference model, and direction of development of the DT, indicating the huge potential of landing applications. In academic and industrial fields, researchers and practitioners have successively proposed some smart industry concepts based on DT\cite{MA2022DTsmartmanufacturing}, including online process monitoring\cite{LI2022SCDT}, product quality control\cite{TURAN2022DTOptim}, and process optimization methods\cite{Flores2021DTsmartproduction}. 

Most DTs are either modeled by expert knowledge or tuned by offline data from production. Erhan et al.\cite{TURAN2022DTOptim} provide a DT modeling application to improve the quality of processes and products. DT is modeled with data from sensors and PLC, finite element simulations, and several data-analytic tools are also combined to minimize material consumption, maximize product performance, and improve process quality. After the DT is developed, a detailed analysis of the process characteristics, process parameters, and process challenges is performed with material modeling to validate the fidelity of the DT. With the help of DT, the thermoforming process is significantly improved and the scrap ratio decreased. However, the validation of the DT still relies on the manual tuning process and expert knowledge. Panagiotis et al. \cite{STAVROPOULOS2022DTPC} describe a method that uses DT to model thermal manufacturing processes. The work proposes DT that generalizes the concept of process control toward multivariable performance optimization. Data-driven models describe the mathematical link between process parameters and performance indicators. Furthermore, the performance of process models is examined by aggregating the details of the modeling from the process data to reduce the gap between theoretical and data-driven models. Flores et al. \cite{Flores2021DTsmartproduction} propose a framework that combines DT-based services and the Industrial Internet of Things for real-time material handling and location optimization. The research shows an improvement in the delivery, makespan and distance traveled during material handling.

DT also shows a great advantage when combined with artificial intelligence (AI) and machine learning (ML). In most applications, DT is driven by real-scene data, online monitoring in real-time, and optimization of the entire process. For example, Jinfeng et al. \cite{LIU2021PQ} describe a DT-driven approach to traceability and dynamic control for processing quality. The research introduced a Bayesian network model to analyze factors that affect processing quality. With the acquisition of real-time data in the IoT system of the manufacturing unit for dynamic control of processing quality, DT-driven dynamic control and management is proposed. Yun-Peng et al.\cite{ZHAO2022DT} describe a DT based on an artificial neural network that detects damage to the fishing net in real-time. The proposed DT can detect the final net damage accurately and quickly, even under complex sea conditions. Industrial robots can be trained in manufacturing to learn dexterous manipulation skills with DRL. However, training industrial robots on real production lines is unrealistic, as the interaction between the DRL agent and the real-world environment is time-consuming, expensive, and unsafe\cite{Meyes2017industrialrobots}. A common approach is to train robots through simulations and then deploy algorithms on physical robots.

The fidelity of DT is significant when combined with other state-of-the-art technologies, for example, by training a DRL agent with DT. To evaluate the fidelity of DT, Shimin et al. \cite{LIU2022DTEVA} construct an adaptive evaluation network for the DT machining system, which evaluation network can predict the decision-making error in the process route and assess its reliability. Houde et al. \cite{SONG2022CALDT} demonstrated an autonomous online calibration of DT using two-stage machine learning. In the offline stage, the simulated data and the corresponding measurement data are used to build an error database. Errors in the database are grouped by clustering to reduce the complexity of the calibration model. The calibration model is continuously updated in the online stage using a dynamic error database while the digital twin runs in parallel with the real operation. Similarly, He et al. \cite{ZHANG2022EVADT} proposed a consistent evaluation framework that uses the analytic hierarchy process to form a comprehensive consistent evaluation. The judgment matrix is constructed in this approach to calculate the consistency ratio. The judgment matrix is updated iteratively to obtain a consistent evaluation framework.

DT-assisted industrial production has achieved significant success in different areas. The combination of DT and state-of-the-art technologies such as AI, IIoT, and ML provides potential industrial digitalization and automation opportunities. However, in the literature, most of the work used DT for industrial optimization, quality testing, online monitoring, and control. Less work has been shown to test the actual DT software online and continuously evaluate the DT performance during production. In this sense, our work shows a data-driven approach to test the DT software. Consistent evaluation of DT is critical for industrial applications, since AI decision-making and other control algorithms closely work with DT, and the fidelity of DT directly impacts the performance of control algorithms centralized with DT.


\section{The Digital Twin Under Test}\label{sec:DTtest}

The DT under test in this paper is used to simulate the induction heating process of the Smart Forge industrial showcase. It should be mentioned that although the DT under test in this paper is specific to the induction heating process of the Smart Forge, our testing method can be applied to other DT applications as long as sensor data is involved. This is because our method focuses on evaluating the accuracy of the sensor data being used by the DT and the robustness, which is a critical aspect of any DT application.

\begin{figure}
    \centering
    \includegraphics[width=0.5\textwidth]{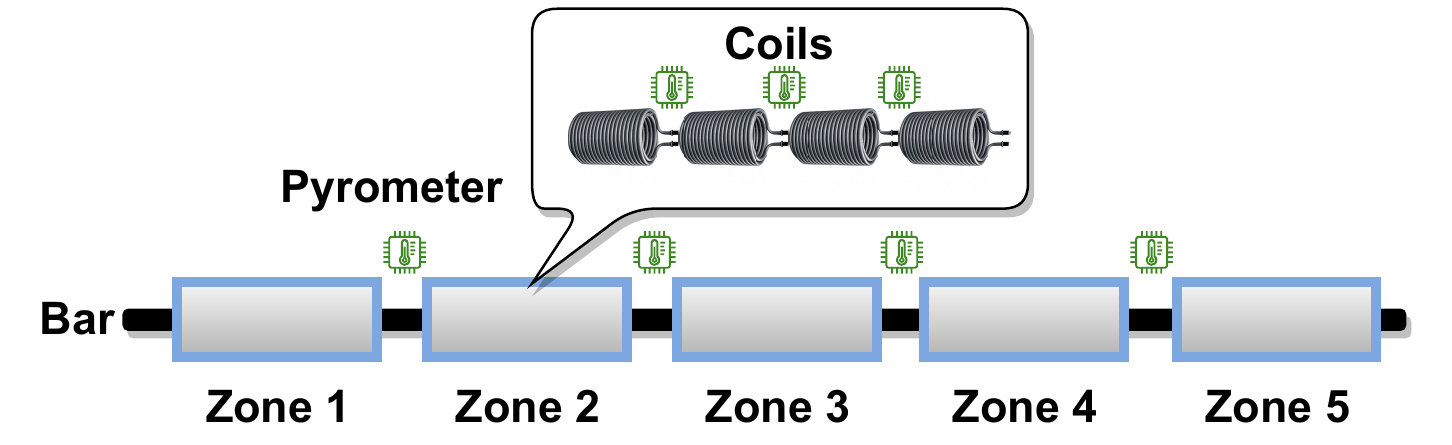}
    \caption{Heating Furnace}
    \label{fig:furnace}
\end{figure}

As shown in Figure \ref{fig:furnace}, the DT is used to simulate the heating process of steel bars passed through the heating furnace. The heating furnace is divided into five zones of electrical induction heaters consisting of four induction coils, each. The temperature between each coil is monitored by pyrometers. In normal production, steel bars move inside the heating furnace toward the shear process outside Zone 5. However, when there is a disturbance downstream of the production line, the furnace is set to holding mode, in which the power to the coils is reduced and the steel bars slowly oscillate back and forth. This is essential to keep the bar temperature at a certain level and not shut down the furnace, as it will take a long time and significant resources to heat it up to a certain temperature again when the disturbance is over. A simulator is developed that aims to represent this production process by incorporating the effects of induction heating, conduction, radiative cooling, and steel bar movement into a DT of the production line. The developed DT can also facilitate our control algorithm design, e.g., combinatorial optimization algorithm and DRL algorithm. In this paper, we also provide the details of our DT-based DRL architecture as shown in Figure \ref{fig:architecture}. The following subsections provide the details of the DT and its components, including our developed ML algorithm to optimize and automate the production line.  

\subsection{The Digital Twin}\label{sec:DT}


\begin{figure*}
    \centering
    \includegraphics[width=0.9\textwidth ]{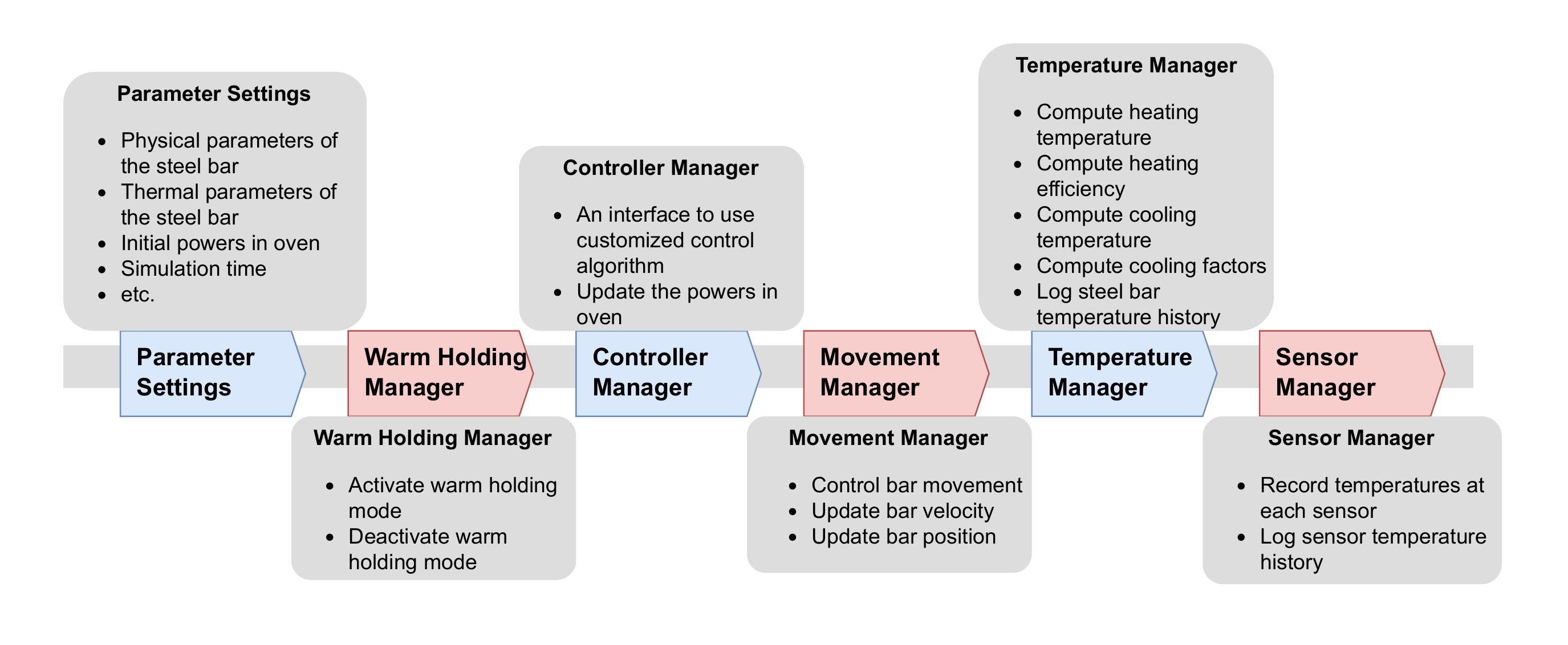}
    \caption{Digital Twin}
    \label{fig:DT}
\end{figure*}


The DT of the forging line simulates the heating process of one or more steel bars by moving them through one or several linearly arranged coils. Each coil is defined by its fixed start and end positions and power settings, which can change dynamically. The DT assumes that only the segment of the steel bar under a coil is heated with the coil power. Other segments of the steel bar are not affected by the coil. The dimensions, mass, and specific heat coefficient define a steel bar. The DT is implemented as a discrete event simulator. The DT simulates two modes:

\textbf{Normal mode:} In normal production, steel bars move at a configurable constant speed towards sheer. The induction coils in the oven are applied with normal production power.

\textbf{Holding mode:} Steel bars move forward and backward with a configurable holding mode speed, and the direction changes at regular intervals. The induction coils in the oven are applied with warm holding power.

The position of the bar along the production line is updated with a \emph{Movement Manager} for each simulation step. The Movement Manager has two modes of operation:

\noindent When the steel bar reaches the end of the heating line, the DT assumes that the bar is moved to the next step of production. Therefore, the DT will remove the bar and record the temperature of the head. The history of movement of the steel bar and its temperature profile for each time step are kept for further analysis. The temperature of the steel bar is updated at each DT time step by a \emph{Temperature Manager}. For every segment along the steel bar (within a configurable resolution), the segment is heated if it is within the coil range. The DT also provides an interface for algorithms to interact with it using the \emph{Controller Manager} class, which continuously provides data to external control algorithms and returns the algorithm responses to the DT. 

As shown in Fig. \ref{fig:DT}, the DT consists mainly of six parts, as follows: 


\begin{enumerate}
    \item \textbf{Parameter settings:} define all parameters in the DT, including physical parameter settings (steel bar's mass, dimension, density, speed, etc.) and thermal parameter settings (specific heat, heating efficiency, cooling factors, etc.) of the steel bar, power settings of each zone, simulation time, coil position, sensor position, etc.
    
    \item \textbf{Warm holding manager:} determines whether the DT simulates normal production mode or warm holding mode.
    
    \item \textbf{Controller manager:} provides an interface to design different control algorithms such that the DT can be used to assist in algorithm development.

    \item \textbf{Movement manager:} updates the speed of the steel bar, which is determined by the operation mode and the properties of the steel bar. The position of the steel bar is also updated at each discrete time step.

    \item \textbf{Temperature manager:} is a crucial part of the DT. Mathematical equations define thermo-heating and cooling models. The temperature manager defines parameters such as the heating efficiency and cooling factors for different materials. Thus, the heating and cooling temperatures of the steel bar can be computed at every discrete time step. The temperature history of the steel bar is also recorded for analysis.

    \item \textbf{Sensor manager:} manages sensor temperature updates. The sensors are located between the induction coils. When the steel bar moves under the sensor, the temperature of the part under the sensor is recorded on the corresponding sensor. The sensor manager also logs all sensor temperature history during simulation time to perform an analysis.

\end{enumerate}


\subsection{DT Assisted Deep Reinforcement Learning Architecture}

\begin{figure}
    \centering
    \includegraphics[width=0.5\textwidth]{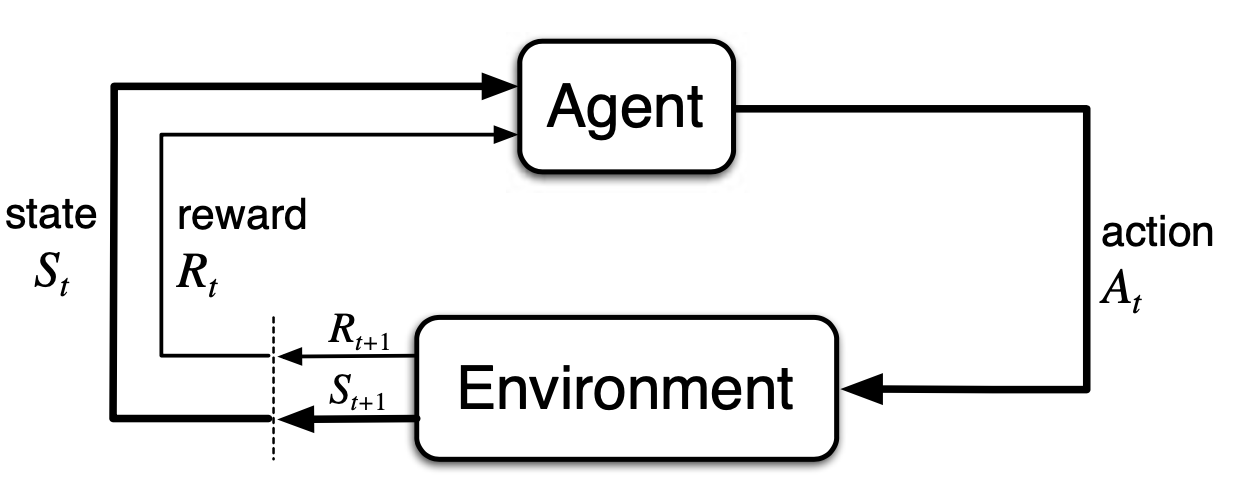}
    \caption{Markov Decision Process}
    \label{fig:mdps}
\end{figure}

Most industrial controllers, such as PID controllers, are model-based, relying on the need for mathematical modeling and knowledge of process dynamics. The performance of classical PID controllers depends only on a well-tuned process model. However, identifying an unstable system model is tedious due to the unbounded nature of the output. Consequently, model-based methods are degraded with unstable process dynamics due to inevitable modeling errors \cite{SHUPRAJHAA2022RLPID}. To address this problem, a data-driven and AI-assisted controller within this DT is used that does not require expert knowledge and can achieve robust control. More specifically, reinforcement learning (RL) is used within the DT to control the production process by predicting and deciding the power to be used in the next steps of the production line. The RL framework shown in Figure \ref{fig:mdps} consists of an agent that interacts with a stochastic environment modeled as a Markov decision process (MDP). The goal of RL is to train an agent that can predict an optimal policy \cite{Spielberg2020DeepRL}. In our application, the agent is the controller of the oven power, and the environment is the DT. Also, in practice, we found that the performance of the DRL will be greatly improved if the DRL agent only controls the power. Hence, we reduce the search dimension by keeping the speed constant, which may be improved by using multi-agent DRL for future work.

\begin{figure*}
    \centering
    \includegraphics[width=0.9\textwidth, ]{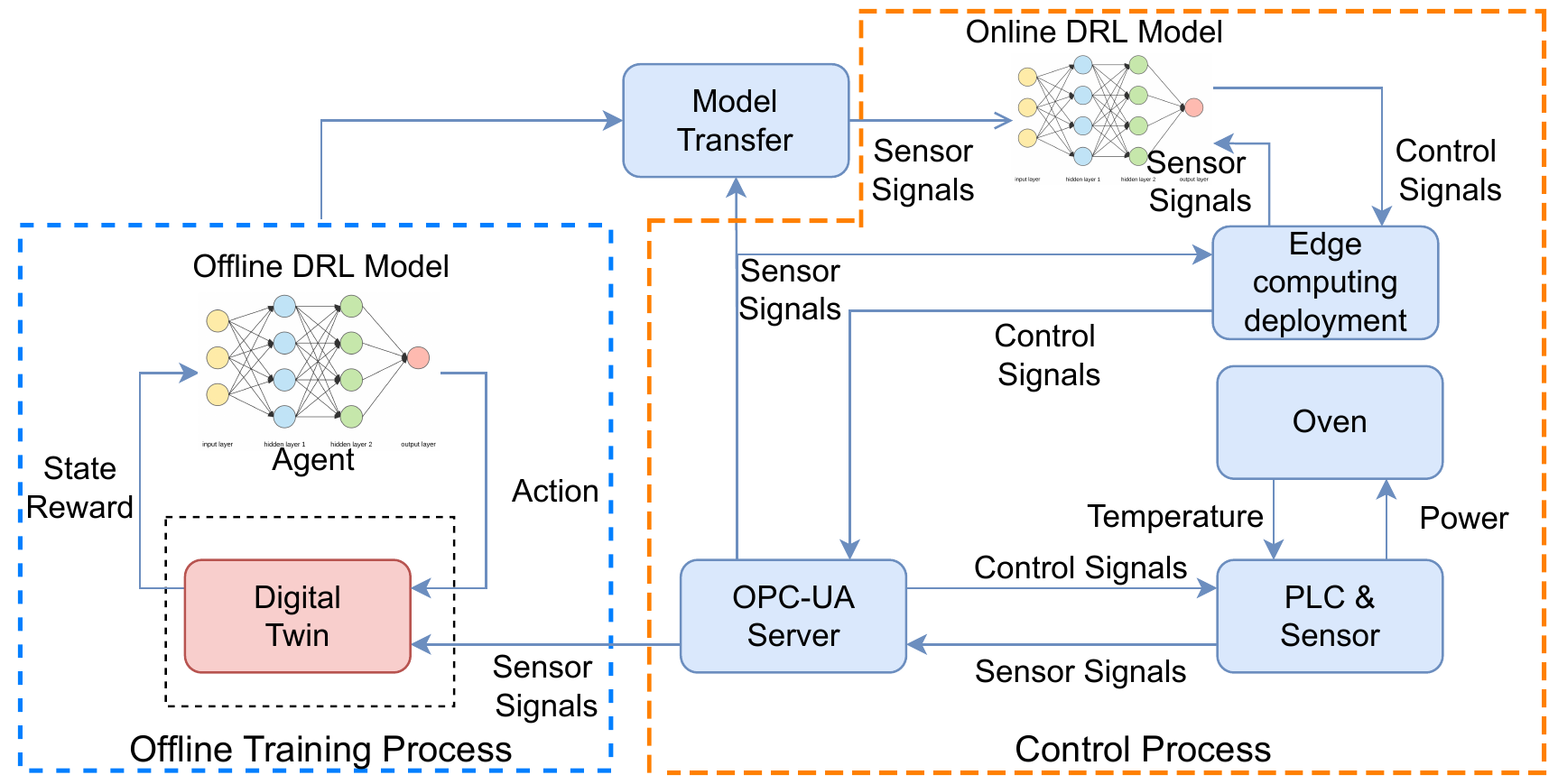}
    \caption{Digital Twin-Based Deep Reinforcement Learning Architecture}
    \label{fig:architecture}
\end{figure*}

Figure \ref{fig:architecture} shows the overall architecture of DT where deep reinforcement learning (DRL) is used within the architecture for decision making. It should be mentioned that the evaluation of the DRL's performance is beyond the scope of this paper. However, the DT serves the decision-making process led by the DRL algorithms. More information on the design and evaluation of this DRL can be found in our latest published study \cite{Ma2022DRL}.

As shown in Figure \ref{fig:architecture}, the simulation model within the DT of the production process acts as an environment for the agent and simulates the physical dynamics at each time step. In the offline training process (blue dashed box), the offline DRL models (holding mode: $DRL\_wh$, normal production: $DRL\_n$) are trained by the agent that observes states and rewards by instrumenting the DT and outputs actions (e.g., adjust power) to the simulation model. The temperatures, positions, and speeds of the steel bars are mapped to the states of the agents. The agent observes the states and immediate rewards to update the neural networks. Once the agent is well trained, it can predict optimal power based on the observed temperatures, positions, and speeds of the steel bars at every time step. Offline models are transferred to online DRL models ($DRL^\ast\_wh$, $DRL^\ast\_n$) using, for example, transfer learning. Online DRL models are deployed on edge compute platform for real-time interaction with the control process to adjust the parameters of the heating process for the forging line (e.g., power levels).

The part in the orange dashed box in Figure \ref{fig:architecture} shows the online control process when using the trained policy. The heating oven is controlled by adjusting the power to the induction coils or changing the speed of the conveyor rollers via a PLC. Actual production data, such as the temperatures in each zone, conveyor movements, and used voltage levels, are recorded through an OPC-UA server. Online DRL models are deployed on an edge computing platform, subscribe to OPC-UA monitoring data, and predict optimal control signals for the next time step, which are forwarded to the PLC by writing to dedicated OPC-UA tags on the server.

As the control policy is trained on a DT, the DT should resemble the conditions of the physical world as much as possible, so the off-line learned policy of the DRL is also a good policy when operating on real production data. If there is a large mismatch between the DT and the real system, the trained control policy will not be a good policy when feeding it with real-time data from the actual production data. Therefore, it is important to have a high-fidelity DT. Automated testing of this DT is essential to ensure high fidelity and accuracy to make robust decisions. By comparing real and simulated data, we can measure the error and use that information to fine-tune the DT after testing.


\section{Real-time DT Testing Architecture} \label{sec: DT_Architecture}

\subsection{Testing Agent Architecture}
\begin{figure*}
    \centering
    \includegraphics[width=0.98\textwidth, ]{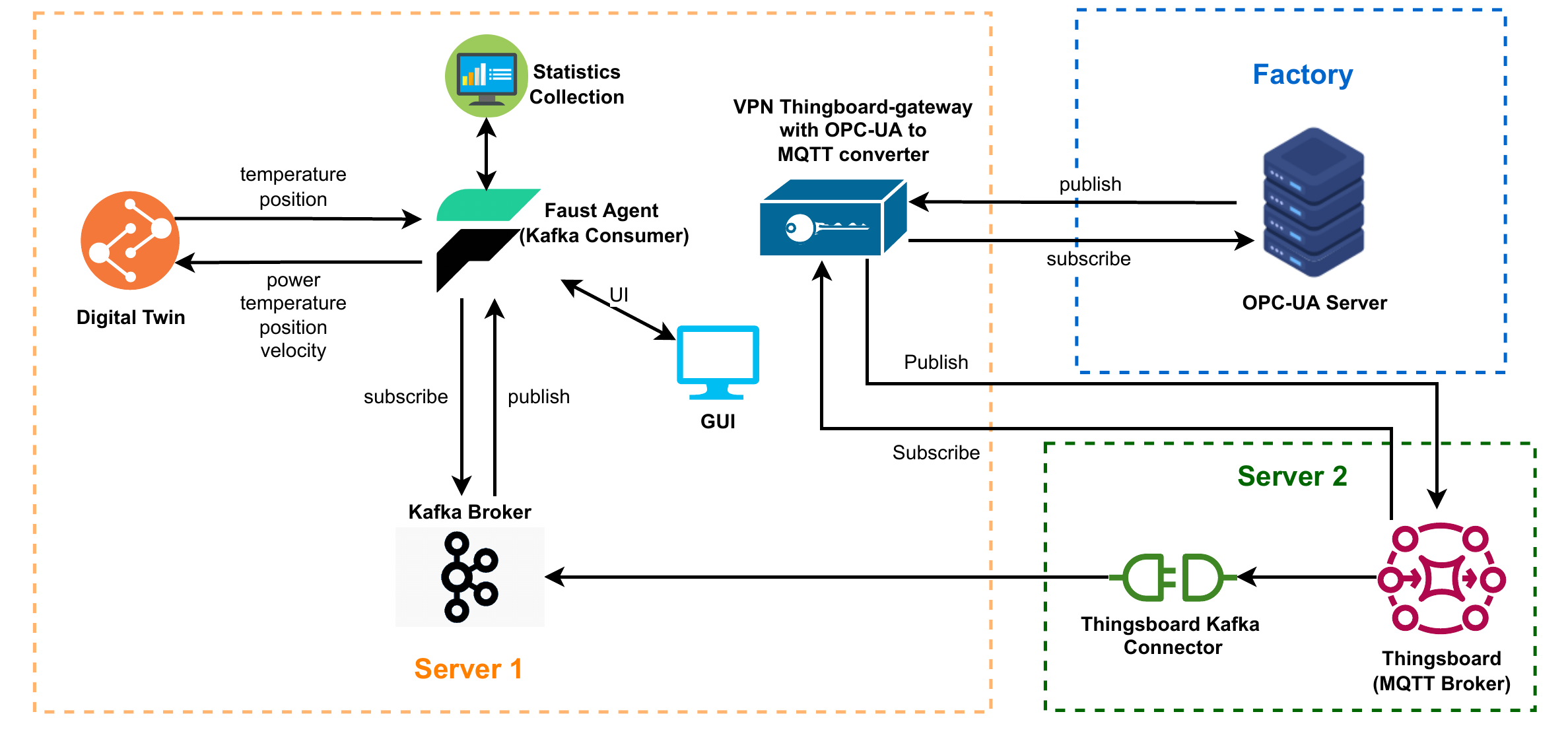}
    \caption{Testing Agent Architecture}
    \label{fig:testingarchitecture}
\end{figure*}

We aim to design and implement an automated testing architecture that can help evaluate the accuracy of DT in an online production setting. The main idea is that the live process data together with the coil power settings enter the DT, which aims to predict the temperature using the DRL, compare the prediction with the real values of the process, and visualize the deviation. Figure \ref{fig:testingarchitecture} shows the architecture of the testing agent. The architecture obtains real data from the OPC-UA server, creates snapshots continuously, and automatically instruments the DT. The DT receives the preceding snapshot from the real data, runs several simulation steps, and compares the output with the oncoming snapshot from the process data. Each snapshot acts as a test case. The deviation calculation acts as a test oracle for each snapshot. The OPC-UA server transfers sensor signal data to other services, including the temperature of the pyrometer sensors, the movement of the conveyor and the voltage levels sent to the induction coils. Thingsboard Gateway\footnote{Thingsboard web page https://thingsboard.io/} subscribes to OPC-UA tags on the factory server and publishes sensor updates as telemetry data to a local Thingsboard instance using the MQTT protocol, which stores and visualizes sensor data. To build high-performance and scalable data pipelines for DT testing, the Thingsboard server uses a Kafka connector to push sensor data to a Kafka broker. We build a Faust\footnote{Faust web page https://faust.readthedocs.io/en/latest/} agent as a Kafka consumer that subscribes to the Kafka broker to provide different functionalities: DT testing, statistics collection, and real-time error visualization through the GUI. The Faust agent instrument the DT by creating snapshots of the current production line in terms of power, temperatures, bar position, and speed, which are sent to the DT. The DT simulates expected new temperatures for the given power settings and bar position, which the Faust agent compares against a new snapshot from the production data. For a perfect DT, the new snapshot from the production data should have a slight error compared to the expected temperature calculated from the DT. 

\subsection{Snapshot Creation Methodology} \label{subsec:snapshot}
The Faust agent continuously receives a data stream with telemetry information that includes the temperature of each sensor, the position of the head and tail of the bar, and the power in each zone. However, the challenge is that the OPC-UA server may lag in time, telemetry may be lost, or the update intervals of each OPC-UA tag may be different. The DT cannot be evaluated if the data are missing or not coherent. Therefore, we propose a snapshot creation algorithm in which we collect snapshots of the data and compare them with the simulation output.

\begin{table}
\centering
\caption{Snapshot Data Structure}
\begin{tabular}{||c | c||}
 \hline
 \textbf{Dataset(1/2)} & \textbf{Values }\\ [0.5ex] 
 \hline\hline
 Power & List[pwr]  \\ 
 \hline
 Temp & List[temp] \\
 \hline
 Position & List[back, head] \\
 \hline
 Speed & Float \\[1ex] 
 \hline
 Timestamp & Float \\[1ex] 
 \hline
\end{tabular}

\label{table:snapshot}
\end{table}

A snapshot is created by continuously collecting telemetry from the real-time stream until the snapshot contains all the information listed in Table \ref{table:snapshot}: the coil power at each zone, the temperature at each sensor for all zones, bar's position including bar head and tail position, speed of the bar, the timestamp. The powers, temperatures, and positions values are logged in a list. The speed is stored as a floating point number. The timestamp, which corresponds to each data sample, is also collected. The main procedure is shown in Algorithm \ref{alg:snapshot}. Steps 4 to 10 explain that we repetitively use Dataset1 as the DT input and compare the DT output with Dataset2. In addition, we also collected other data to facilitate our analysis, such as the hold mode indicator, which can help to differentiate data on the normal production mode or hold mode. This indicator can also be verified by the speed moving direction. However, they do not directly contribute to our snapshot creation algorithm.

\begin{equation}\label{equa: simulation time}
    t_{simulation} = (Position_{Dataset2} - Position_{Dataset1})/Speed
\end{equation}

\alglanguage{pseudocode}
\begin{algorithm}
\scriptsize 
\caption{Snapshot Creation Algorithm}
\label{alg:snapshot}
\begin{algorithmic}[1]
\State $Dataset1 \gets Snapshot1$
\State $Dataset2 \gets Snapshot2$
\While {True}
\State Use Dataset1 as input for the DT
\State Calculate simulation time $t_{simulation}$by Equation \ref{equa: simulation time}
\State Run the DT for $t_{simulation}$ time steps
\State Compare DT output with Dataset2
\State Copy Dataset2 to Dataset1, empty the position in Dataset2
\State Overwrite temperature, power, and speed in Dataset2 until getting a new position
\State Collect snapshot and store it in Dataset2
\EndWhile

\end{algorithmic}
\end{algorithm}


\section{Experimental Evaluation} \label{sec:results}
Our experimental evaluation aims to understand whether the DT can be automatically tested with the online data stream. By analyzing the test results, we aim to know whether the movement pattern, position update, and sensor temperature update in the DT align with real production. The details of the DT structure are mentioned in Section \ref{sec:DT}. The parameter setting module, warm holding manager, and controller manager receive inputs that include parameter settings, whether the operation is in normal production or warm holding mode, and power settings to the DT. They do not contribute to the fidelity of the DT. In addition, the temperature manager defines the internal temperature update over the steel bar, which cannot be observed in real production. Hence, it cannot be tested directly. The fidelity of the DT is directly reflected by the movement manager, which updates the steel bar's speed and steel bar position, and the sensor manager, which records the steel bar's temperature. We analyze the movement pattern and position error of the steel bar to evaluate the movement manager and the sensor temperature error matrix to evaluate the sensor manager, which can indirectly reflect faults in the DT temperature manager. 

\subsection{Testing Setup}

The testing data come from the OPC-UA server located at Bharat Forge Factory. The DT is developed by VikingAnalytics and is deployed on the server at Karlstad University, together with the Kafka broker and the Faust Agent (Kafka consumer). The Thingsboard Gateway subscribes to OPC-UA tags and publishes data stream to a local Thingsboard instance through MQTT protocol, the Kafka connector pushes the data stream to the Kafka broker, and the Faust agent as a Kafka consumer instrument the DT by creating snapshots and compares the DT output with the new snapshot from the production data. As we mentioned earlier, each snapshot is considered a test case, and the deviation between the two snapshots of the same instance is a test oracle for each test case. Here, if the deviation is more than the deviation specified in the oracle, the test case is considered failed. The fault will be localized within the snapshot data.


\subsection{Testing Results}
\begin{figure*}
    \centering
    \subfloat[Steel bar's movement]{\includegraphics[scale=0.4]{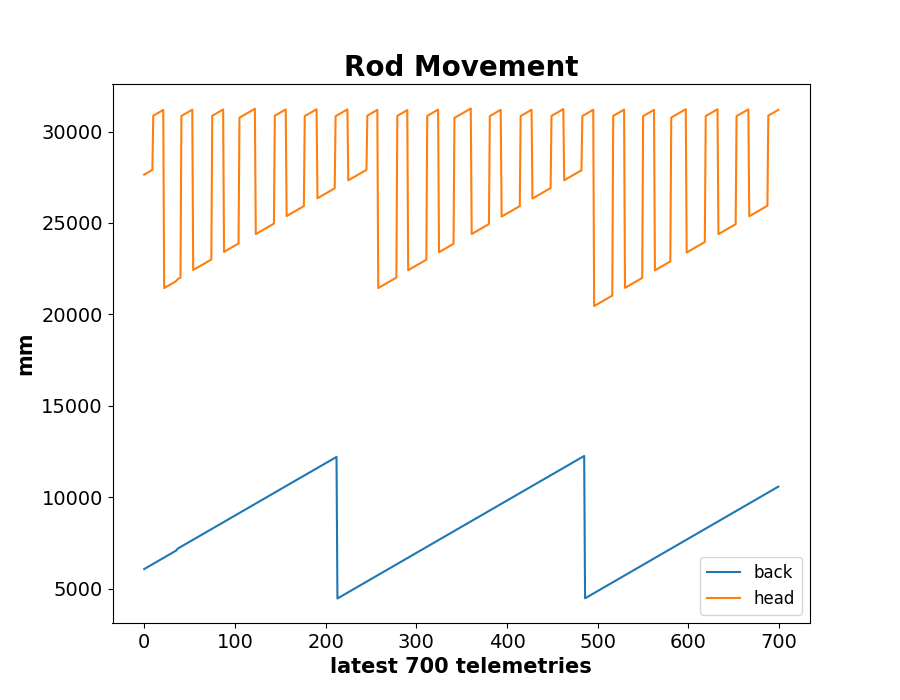}}
     \subfloat[Sensor\_Position: Position error]{\includegraphics[scale=0.4]{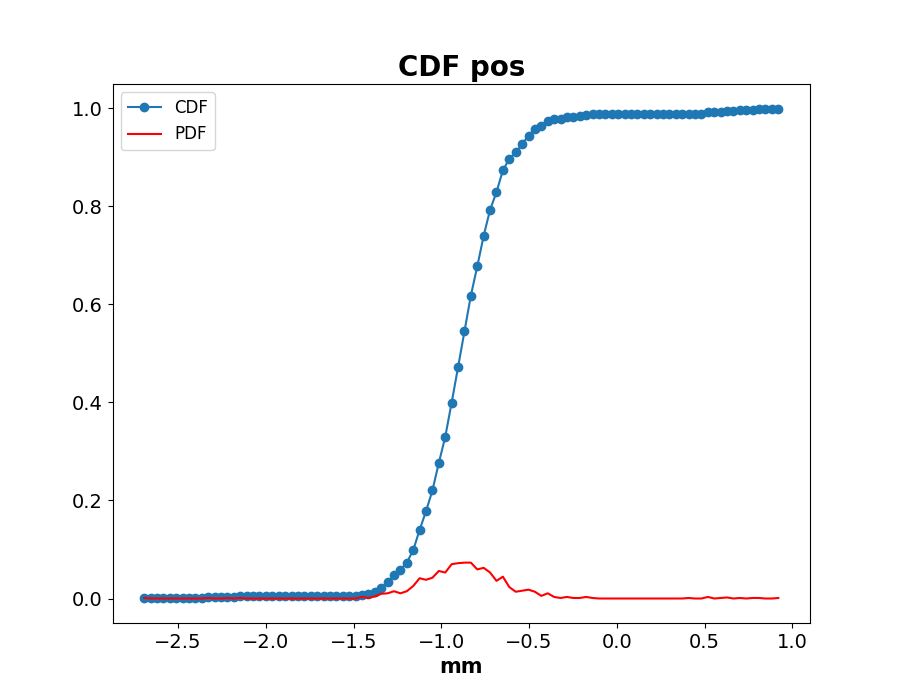}}

    \subfloat[Sensor1\_3: Temperature error]{\includegraphics[scale=0.4]{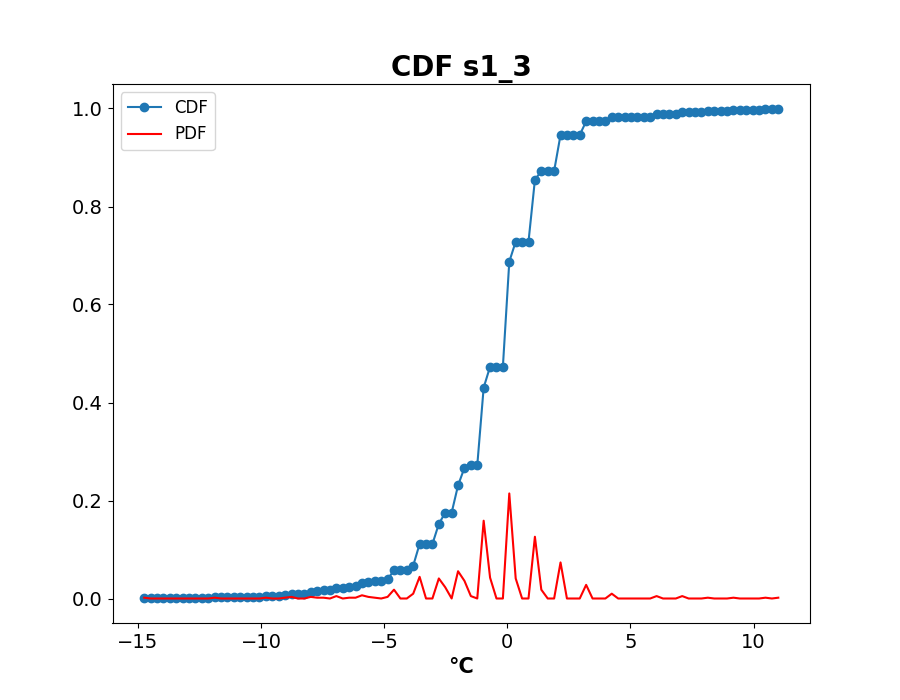}}
    \subfloat[Sensor2\_2: Temperature error]{\includegraphics[scale=0.4]{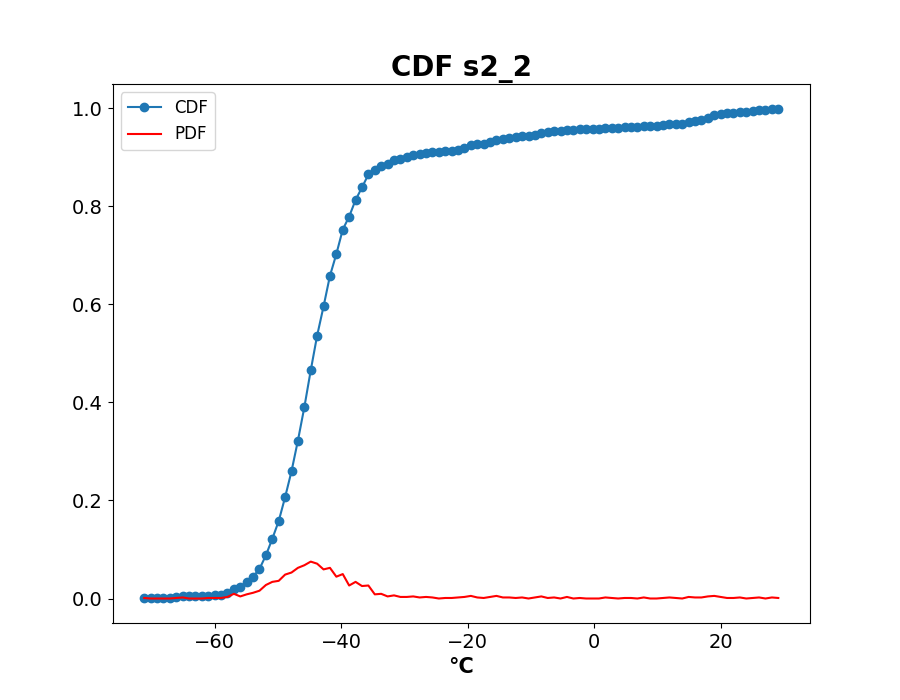}} \\
    \caption{(a). Shows the movement of the tail and head of the steel bar during normal production mode. Probability density function(PDF) and cumulative distribution function(CDF) of errors for each sensor in this figure: (b). The position sensor (c). The $3^{rd}$ sensor in Zone 1 (d). The $2^{nd}$ sensor in Zone 2}
    \label{fig:resultNP}
\end{figure*}

\begin{figure*}
    \centering
    \subfloat[Steel bar's movement]{\includegraphics[scale=0.4]{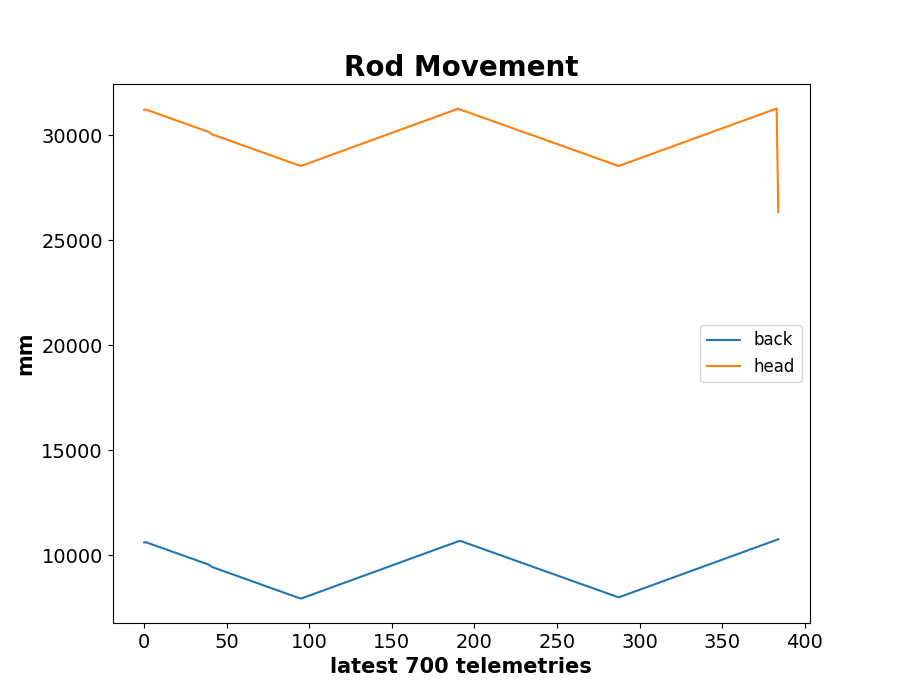}}
     \subfloat[Sensor\_Position: Position error]{\includegraphics[scale=0.4]{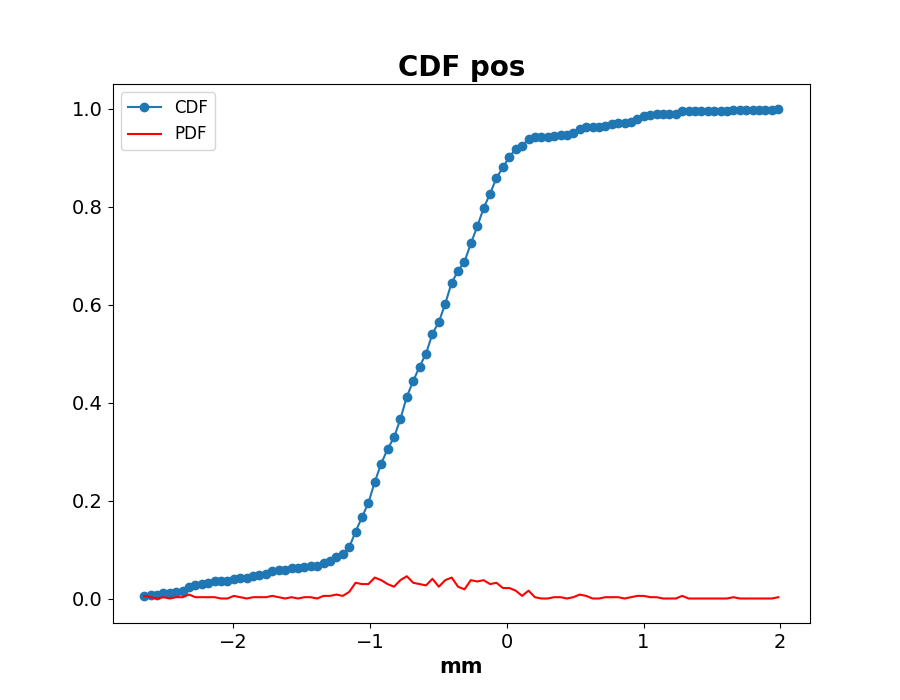}}

    \subfloat[Sensor1\_3: Temperature error]{\includegraphics[scale=0.4]{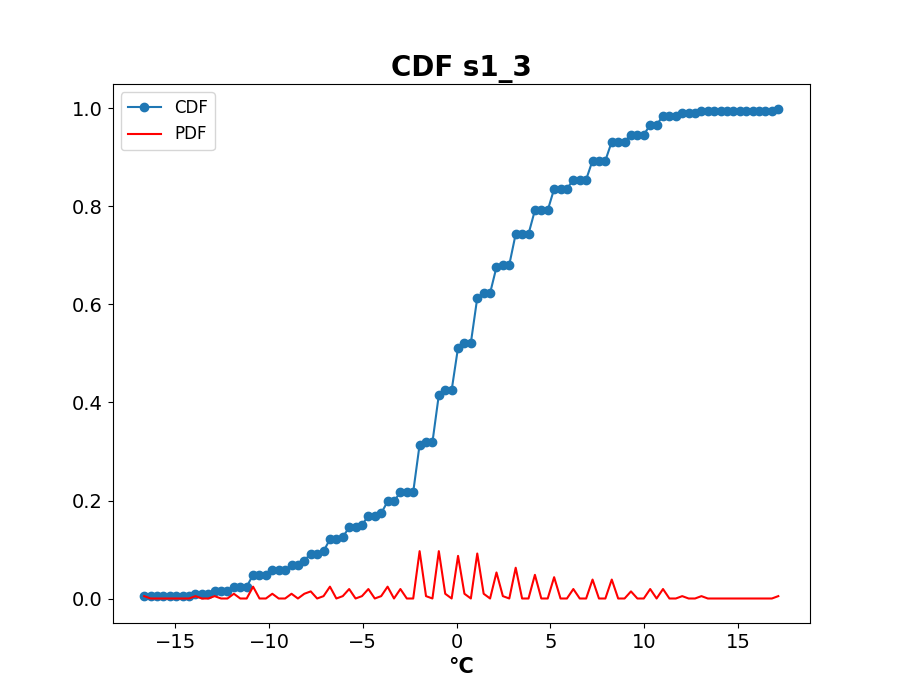}}
    \subfloat[Sensor2\_2: Temperature error]{\includegraphics[scale=0.4]{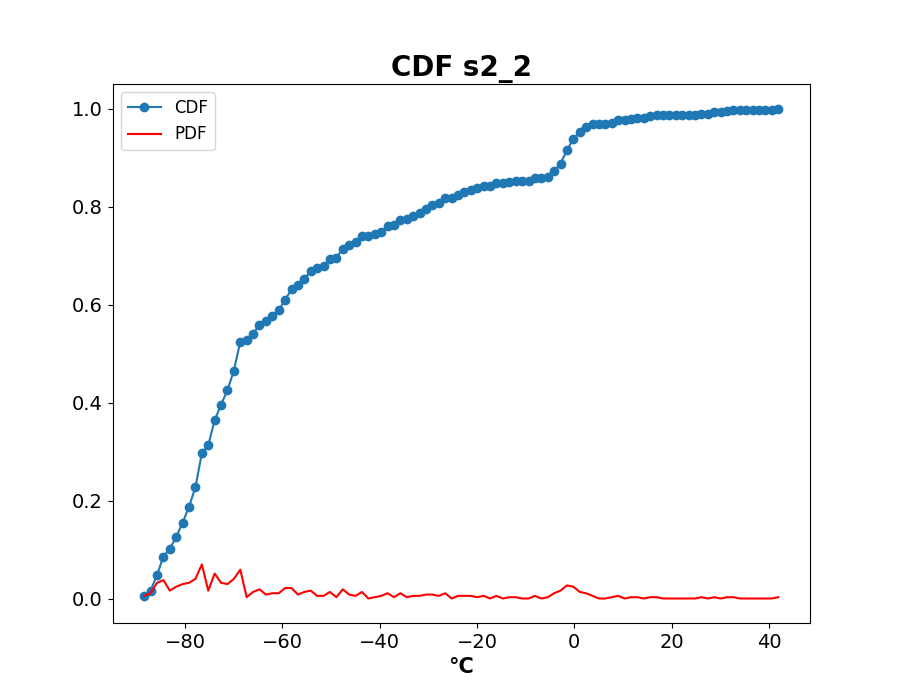}} \\
    \caption{(a). Presents the movement of the tail and head of the steel bar during holding mode. Probability density function(PDF) and cumulative distribution function(CDF) of errors for each sensor in this figure: (b). The position sensor (c). The $3^{rd}$ sensor in Zone 1 (d). The $2^{nd}$ sensor in Zone 2}
    \label{fig:resultWH}

\end{figure*}

Figure \ref{fig:resultNP}(a) shows the movement of the steel bars in the furnace during normal production. The front bar of the consecutive rods leaves the furnace and enters a cutting-sequence cycle because the head moves back and forth. The position of the tail is dropped at telemetry 200 and 500 due to new bars being added to the furnace. During the hold mode, the consecutive steel bars move forward and back at a constant speed. This can clearly show whether the current operation is on normal production mode or hold mode, which corresponds to the hold mode indicator. Figure \ref{fig:resultWH}(a) shows how the steel bars move and change direction at regular intervals. We continuously run the testing on production data. The test oracle calculates the error for each snapshot and sensor within a snapshot, which is the difference between the estimated temperature and position of DT and the temperature and position of the real production data, and decides which snapshot (i.e., test case) passed and failed. We present results from the position sensor and two temperature sensors (one in Zone 1 and one in Zone 2).

As can be seen, the position sensor in Figure \ref{fig:resultNP}(b) shows a position error of -2.5mm to 1.0mm with a spike in the probability distribution function (PDF) around -0.8mm during normal production that indicates a fault in the DT. The error deviation during the hold mode (Figure \ref{fig:resultWH}(b)) is between -3 and 2mm with a spike between -1 and 0mm. Temperature sensor Sensor1\_3 in Zone 1 deviates during normal production from the expected one by the DT between -15°C to 10°C as shown in Figure \ref{fig:resultNP}(c) with a PDF spike at 0°C (no deviation). According to the cumulative distribution function (CDF) graph, the probability that the DT deviates between -4°C and 5°C is approximately 95-10=85\%. However, during the holding mode (Figure \ref{fig:resultWH}(c)), the deviation of the DT increases, and the probability of deviation between -4°C and 5°C during the holding mode is 85-20=65\%.
The second temperature sensor in zone 2 (Sensor2\_2) deviates during normal production from the expected value calculated by the DT within the interval -70°C to 30°C (Figure \ref{fig:resultNP}(d)). In the holding mode, the deviation is between -90°C and 40°C (Figure \ref{fig:resultWH}(d)). The PDF in Figure \ref{fig:resultNP}(d) shows an error spike at around -45°C during normal production. However, during holding mode, errors are mainly distributed around -80° C and 0° C (Figure \ref{fig:resultWH} (d)).

\subsection{Discussion}

As can be seen in the results, the temperature deviation between the DT and the real production happened a few times, indicating a fault in the DT software. Without the continuous testing process and architecture, detecting and fixing those faults was impossible. Apart from those fixed faults, there are also multiple reasons for the significant temperature deviation between the DT and the real production. During the time between the collection of snapshots, the coils' power may have been updated. Considering that the DT does not update the power during this time period, the temperature output could differ in such cases, especially if a large power change occurred. For example, if the power for zone 3 is at high power in Dataset1 but suddenly changes to low power before Dataset2 is collected, the DT would run the coils' power at high power for zone 3 during the whole process, which may lead to temperature errors as reported by the DT. Another reason could be that some sensors are not calibrated well in reality and constantly record the temperature with a significant deviation, while it does not exist in the DT. Furthermore, the current DT does not consider radial or axial thermal conduction inside the steel bar, which, in reality, is especially prominent at the start of the heating process when there is a large temperature gradient between the surface and the core. Temperature sensors only record the surface temperature. All of these approximations affect the experimental results. However, in this paper, our aim is to provide a systematic DT testing method that can improve other researchers or practitioners to speed up their testing using the data-driven approach.

\section{Threats to Validity and Generalizability of the Implementation}\label{ThreatsToValidity}

There are several threats to the validity of our study on automated and systematic digital twins testing for industrial processes, and here are some ways we tried to mitigate them. One threat is selection bias, as we only included samples from our industrial processes case study. To mitigate this threat, we used a random sampling approach to select the processes for our study and ensured that our sample was diverse in terms of size, duration, and complexity. Furthermore, we continuously executed our testing process in the deployed architecture for a long time to capture snapshots. 

Another threat is constructed validity, as we used a self-developed testing approach to evaluate the DT. To mitigate this threat, we thoroughly reviewed the literature to ensure that our approach was grounded in existing theories and practices. We sought feedback from industry experts in the field to validate the suitability of our approach. Furthermore, there is a threat of measurement error, as we relied on various sensors and monitoring devices to collect data on the DT. To mitigate this threat, we carefully calibrated all our devices before the study and performed ongoing checks throughout the study to ensure their accuracy. Overall, these efforts helped to reduce the potential impact of these threats on the validity of our study.

In terms of the generalizability of the proposed testing method, we believe it will likely apply to a wide range of industrial processes that use DTs. This is because our approach is systematic and automated, allowing it to be easily reproduced and applied to other contexts. In addition, we designed our approach to be flexible and adaptable so that it can be tailored to the specific needs of different processes. To further support the generalizability of our findings, we plan to conduct additional studies in different industrial settings and with different types of DT to examine the robustness of our approach. It is important to note that while the DT we are testing in this paper is specifically designed for the induction heating process at Smart Forge, our testing method can be utilized for other DT applications as long as they utilize sensor data. This is because our method evaluates the accuracy of the sensor data being used by the DT and its simulated data generation, a crucial aspect of any DT application. By verifying the accuracy of the sensor data, we can increase the overall accuracy and effectiveness of the DT. Therefore, our testing method can be applied to various DT applications in different fields.

\section{Conclusion and Future Work} \label{sec:conclusion}

This paper introduced a systematic testing architecture with production data that aims to achieve automated DT testing. We presented a systematic method using the DT to train a DRL model and how the DT simulates the induction heating line. We proposed our snapshot creation algorithm to test the DT with real production data in a real-time setting. The evaluation results show that our approach can automatically detect faults in the DT by using real production data, such as faults in the movement of the steel bar. Our method also analyzes error metrics of the temperature and position of the steel bar. Future work will focus on improving snapshot creation algorithms considering corner cases where sudden power changes occur during the DT run time, and improving DT accuracy by considering radial or axial thermal conduction inside the bar.

\section*{Acknowledgement}
This work was partially funded by Vinnova through the SmartForge project. Additional funding was provided by the Knowledge Foundation of Sweden (KKS) through the Synergy Project AIDA - A Holistic AI-driven Networking and Processing Framework for Industrial IoT (Rek:20200067).

\vspace{12pt}
\bibliographystyle{IEEEtran}
\balance
\bibliography{my_reference}

\end{document}